\documentclass[a4paper,12pt]{article}
\usepackage{ifpdf}

\usepackage[usenames,dvipsnames]{xcolor}
\usepackage{hyperref}
\usepackage{mathrsfs}
\usepackage{amsmath}
\usepackage{amssymb}
\usepackage{graphicx}
\usepackage{color}
\definecolor{darkblue}{cmyk}{0.9,0.9,0,0}
\definecolor{darkgreen}{rgb}{0,0.55,0}

\usepackage[font=small,labelfont=bf]{caption}

\usepackage{wrapfig}

\definecolor{darkblue}{cmyk}{0.9,0.9,0,0}
\definecolor{darkgreen}{rgb}{0,0.55,0}

\usepackage{epsfig}
\usepackage{amsmath}

\newcommand{\comment}[1]{}

\newcommand{\beq}{\begin{equation}}
\newcommand{\eeq}{\end{equation}}
\newcommand{\beqq}{\begin{equation*}}
\newcommand{\eeqq}{\end{equation*}}
\newcommand\beqa{\begin{eqnarray}}
\newcommand\eeqa{\end{eqnarray}}
\newcommand\beqaa{\begin{eqnarray*}}
\newcommand\eeqaa{\end{eqnarray*}}
\newcommand\bea{\begin{array}}
\newcommand\eea{\end{array}}

\def\XXint#1#2#3{{\setbox0=\hbox{$#1{#2#3}{\int}$ }
\vcenter{\hbox{$#2#3$ }}\kern-.5\wd0}}

\def\XXint#1#2#3{{\setbox0=\hbox{$#1{#2#3}{\int}$}
\vcenter{\hbox{$#2#3$}}\kern-.5\wd0}}

\newcommand{\nn}{\nonumber}

\newcommand{\neqa}{\nonumber\end{eqnarray}}

\newcommand{\eq}[1]{(\ref{#1})}

\newcommand{\Tr}{{\rm Tr}}

\newcommand{\hs}{\frac{\sqrt{3}}{2}}
\renewcommand{\d}{\partial}

\newcommand{\<}{{\langle}}
\renewcommand{\>}{{\rangle}}

\newcommand{\re}{\relax{\rm I\kern-.18em R}}

\renewcommand{\sp}{p\hspace{-.40em}/}

\def\su2{{SU(2)}}

\def\[{\left[}
\def\]{\right]}

\def\s{\sigma}

\def\({\left(}
\def\){\right)}
\def\[{\left[}
\def\]{\right]}

\def\<{\langle}
\def\>{\rangle}

\def\cO{{\cal O}}

\def\mC{{\mathbb C}}

\def\s*{\ *_{\!\!\!\!\!\!\!\!\!\,_{\,_\text{\scriptsize{sym}}}}}
\def\hs*{\ \hat{*}_{\!\!\!\!\!\!\!\!\!\,_{\,_\text{\scriptsize{sym}}}}}
\def\d{\partial}

\def\i2{\frac{i}{2}}

\def\spi{\relax{\rm \pi\kern-0.5em /}}
\def\sA{\relax{\rm A\kern-0.5em /}}
\def\sp{\relax{\rm p\kern-0.5em /}}
\def\sd{\relax{\rm \d\kern-0.5em /}}
\def\sk{\relax{\rm k\kern-0.5em /}}
\def\sn{\relax{\rm n\kern-0.5em /}}
\def\sl{\relax{\rm l\kern-0.5em /}}
\def\sP{\relax{\rm P\kern-0.7em /}}
\def\sBethe{\relax{\rm \Bethe\kern-0.5em /}}
\def\cN{{\cal N}}

	\newcommand{\ofrac}[1]{\frac{1}{#1}}

\newcommand{\mR}{\mathbb{R}}
\newcommand{\mZ}{\mathbb{Z}}

\newcommand{\bra}[1]{\langle #1 |}
\newcommand{\ket}[1]{| #1 \rangle}

\newcommand{\cP}{\mathcal{P}}
\newcommand{\cI}{\mathcal{I}}

\newcommand{\upa}{\uparrow}
\newcommand{\dna}{\downarrow}

\newcommand{\dint}{-\!\!\!\!\!\!\!\!\;\int}

\usepackage[english]{babel}

\numberwithin{equation}{section}

\begin{document}

%
%
%
%
%

\begin{titlepage}

\setcounter{page}{0}
\renewcommand{\thefootnote}{\fnsymbol{footnote}}


\vspace{1cm}

\begin{center}

\textbf{\large\mathversion{bold} Lectures on the Bethe Ansatz}

\vspace{1cm}

{\large Fedor Levkovich-Maslyuk \footnote{{\it E-mail:\/}
{\ttfamily fedor.levkovich@gmail.com}}
} 

\vspace{1cm}

\it Mathematics Department, King's College London\\
The Strand, WC2R 2LS London, United Kingdom

\vspace{1cm}

{\bf Abstract}
\end{center}
\vspace{-.3cm}
\begin{quote}
We give a pedagogical introduction to the Bethe ansatz techniques in integrable QFTs and spin chains. We first discuss and motivate the general framework of asymptotic Bethe ansatz for the spectrum of integrable QFTs in large volume, based on the exact S-matrix. Then we illustrate this method in several concrete theories. The first case we study is the $SU(2)$ chiral Gross-Neveu model. We derive the Bethe equations via algebraic Bethe ansatz, solving in the process the Heisenberg XXX spin chain. We discuss this famous spin chain model in some detail, covering in particular the coordinate Bethe ansatz, some properties of Bethe states, and the classical scaling limit leading to finite-gap equations. Then we proceed to the more involved $SU(3)$ chiral Gross-Neveu model and derive the Bethe equations using nested algebraic Bethe ansatz to solve the arising $SU(3)$ spin chain. Finally we show how a method similar to the Bethe ansatz works in a completley different setting, namely for the 1d oscillator in quantum mechanics. 
\\

This article is part of a collection of introductory reviews originating from lectures given at the YRIS summer school in Durham during July 2015.

\vfill
\noindent 

\end{quote}

\setcounter{footnote}{0}\renewcommand{\thefootnote}{\arabic{thefootnote}}

\end{titlepage}

\tableofcontents

\newpage 

\section{Introduction}

Integrable QFTs in 1+1 dimensions have a vast array of remarkable properties. In particular, it is often possible to exactly compute the S-matrix of the theory (an extensive discussion of this aspect is given in another part of the present collection \cite{Smcourse}). The S-matrix captures a lot of the theory's dynamics but only describes the scattering of {\textit{asymptotic}} states, i.e. particles starting from infinitely far away and then flying off to infinity again. In contrast, another interesting setup to consider is when our theory is put into a spatial box of {\textit{finite}} size $L$. In finite volume the spectrum of the Hamiltonian becomes discrete, so a natural question to ask is what are the energies of the states. 


It turns out that for integrable theories this energy spectrum can be computed to a large extent using only the scattering data. For large volume $L$ one can write down equations for the spectrum in terms of the exact S-matrix. These equations are known as the asymptotic Bethe ansatz equations and they will be the main topic of this article. Like the S-matrix itself, they are exact at any value of the coupling constants. These equations are only valid when $L$ is large  (in a sense that will be made more precise later), but still provide a lot of important information. They are also the first step towards formulating the so-called Thermodynamic Bethe ansatz (TBA) equations which give the energies exactly at {\textit{any}} $L$ including all corrections. The TBA approach is covered in detail in a different part of this collection \cite{TBAcourse} (see also the introductory article of this collection). 

Importantly, both the asymptotic Bethe ansatz and the TBA have been crucial for the recent applications of integrability to several AdS/CFT dualities between gauge and string theories \cite{Beisert:2010jr}. A key problem in this setting is computing the energies of multiparticle string states in finite volume, which are mapped to operator conformal dimensions in gauge theory (in fact the volume $L$ corresponds to the number of elementary fields in the operator). Integrability methods have led to great success in exploring this problem, and in particular in the computation of superstring energies on the $AdS_5 \times S^5$ space which coincide with operator dimensions in the dual $\cN=4$ supersymmetric Yang-Mills theory in four dimensions.
We hope that several simpler examples discussed in this article will serve as a starting ground for understanding how Bethe Ansatz works in the AdS/CFT context. 


The name 'Bethe ansatz equations' originates from the famous solution of the XXX spin chain by Hans Bethe in \cite{Bethe:1931hc}. While our main goal is to study integrable QFTs we will see that often computing the spectrum of some QFT model leads to an auxiliary spin chain which should be solved first. We will see several examples and discuss the Bethe ansatz solutions of these spin chains as well. At the same time, various spin chain models are interesting on their own as many of them find important applications in condensed matter physics.

We will first discuss the asymptotic Bethe equations in a general setting and then cover several examples for particular models.
The presentation is structured as follows. In section 2 we give physical motivation for the asymptotic Bethe ansatz in integrable QFT and write the Bethe equations in a generic form as a periodicity condition on the wavefunction. In section 3 we present in a general form the algebraic Bethe ansatz approach allowing to greatly simplify the periodicity constraint by reducing it to a transparent diagonalization problem. In section 4 we demonstrate the method in action on the example of the $SU(2)$ chiral Gross-Neveu model. In the process we obtain the solution of the celebrated XXX $SU(2)$ Heisenberg spin chain. In section 5 we discuss some features of the XXX chain and its Bethe eigenstates, as well as the classical limit leading to finite-gap equations. In section 6 we proceed to the more complicated case of the $SU(3)$ chiral Gross-Neveu model which we solve via {\textit{nested}} algebraic Bethe ansatz. Finally in section 7 we illustrate the versatility of Bethe ansatz by applying a Bethe-like method to solve the 1d quantum mechanical oscillator. Some exercises for the interested reader are also included throughout the text.

There is certainly a large literature on the subject available, in particular we would like to point out several reviews discussing various aspects of the Bethe ansatz methods \cite{Faddeev,Sutherland,Karbach,Hubbook,Volin,Stijn}. For reasons of presentation clarity, only some selected references are included in this pedagogical article\footnote{We also tried to make the notation in this article maximally consistent with other parts of the present review volume.}.

\section{Asymptotic Bethe ansatz equations in 2d integrable QFTs}

In this section we will discuss the Bethe equations for the spectrum of a generic 1+1 dimensional integrable theory. We will always consider a theory on a circle\footnote{By a circle we actually mean a straight line segment $[0,L]$ whose endpoints $x=0$ and $x=L$ are identified.} of length $L$, i.e. we impose periodic boundary conditions.

As discussed in the part of this collection of articles dedicated to integrable S-matrices \cite{Smcourse}, in an integrable theory the scattering has several remarkable features:
\begin{itemize}
	\item The number of particles is conserved in any scattering event
	\item The momenta do not change but can only be redistributed between particles
	\item The S-matrix for multiparticle scattering factorizes, i.e. the S-matrix for any number of particles is a product of the $2\to 2$ S-matrices
\end{itemize}

Although in general one cannot introduce a wavefunction in QFT due to the production of virtual particles, for integrable theories these special features (most importantly the first one) make it possible to do this at least in some regions of the configuration space. Then from the periodicity of the wavefunction one can derive quantization conditions which determine the spectrum. 

Let us first discuss a toy model -- a theory with only one particle in its spectrum. An example is the sinh-Gordon model for some values of the parameters. Then an intuitive picture which gives the correct equations for the spectrum is the following one (for a more rigorous discussion see e.g. \cite{Zamolodchikov}). Since the number of particles is conserved we can speak of a wavefunction as in quantum mechanics. If we have $n$ particles on a circle the wavefunction must be periodic. Imagine that we take the first particle around the circle once, eventually bringing it back to its place again. 

\begin{figure}
\begin{center}
\includegraphics[natwidth=384,natheight=253,scale=0.8]{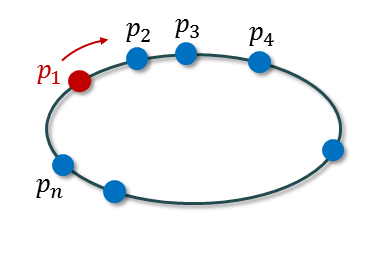}
\end{center}
\caption{\textbf{Deriving the periodicity condition on a circle.} We take the first particle with momentum $p_1$ around the circle, scattering it through all the other particles.}
\end{figure}

If there were no other particles (or if the theory was non-interacting) the wavefunction would acquire a phase factor $e^{i p_1 L}$ where $p_1$ is the particle's momentum. Then from periodicity of the wavefunction
\beq
	e^{i p_1 L}=1\ ,
\eeq	
	i.e. the momentum would be quantized according to
\beq
	p_1=\frac{2\pi k}{L},\ \ k \in {\mathbb{Z}} \ \ .
\eeq
However we need to take into account the interaction with other particles. If $L$ is large compared to the interaction range between particles (e.g. the inverse mass scale of the theory), the particles are almost always well separated from each other. Because of this their interaction is described by the asymptotic S-matrix which we know. The number of particles does not change in this interaction. Thus when we take a particle around the circle, it will scatter through all the other particles (see Fig. 1), and all that happens is that for each scattering the wavefunction is multiplied by the S-matrix which is just some phase factor, $S(p_1,p_2)=e^{i \alpha(p_1,p_2)}$. This product of S-matrix phases will combine with the $e^{i p_1 L}$ phase acquired due to free propagation. Thus periodicity of the wavefunction will be ensured if
\beq
	e^{ip_1 L}S(p_1,p_2)S(p_1,p_3)\dots S(p_1,p_n)=1
\eeq
Similarly for any particle we get
\beq
\label{aba1part}
	e^{i p_j L}\prod_{k=1,\;k\neq j}^n S(p_j,p_k)=1
\eeq
This is a set of $n$ algebraic equations for $n$ variables $p_1\dots,p_n$ which thus allow us to fix the values of the momenta. The system \eq{aba1part} are the asymptotic Bethe ansatz equations for our toy model.

Once the momenta are found from \eq{aba1part}, one can compute the energy of the state. In regions where the particles are well separated they propagate freely, so the energy of the state should always be equal to simply the sum of individual particles' energies. If the energy of a single particle with momentum $p$ is $\epsilon(p)$ we will have
\beq
	E=\sum_i \epsilon(p_i).
\eeq


From the derivation it is clear that the asymptotic Bethe equations will describe the energy only in a large volume $L$. It turns out that, more precisely, the Bethe equations capture all terms in the large $L$ expansion of the energy which scale as $\sim 1/L^k$ (with integer $k$), i.e. the powerlike corrections.
However they miss the exponential corrections of the kind $\sim e^{-mL}$ (where $m$ is the particle's mass) which physically correspond to the effects of virtual particles propagating around the circle.

In a more general theory one could have different types of particles, and different types could transform into each other during scattering.
In this case the S-matrix would have some matrix structure with indices labeling the incoming and the outgoing particles. Then in \eq{aba1part} the product in the r.h.s. would actually be a product of matrices, and should be understood as acting on a wavefunction which also carries indices corresponding to particle types.

Let us first present without derivation how Bethe equations will look like in this case. We will need to introduce some important notation. Let us consider a model with $K$ possible particle flavours, or in other words $K$ particle types. For each particle we should also allow linear combinations of different flavours so its flavour state can be though of as an element of $\mC^K$.  For $n$ particles, their state which we denote by $A$ is then the element of a tensor product 
\beq
	A \in H_1 \otimes H_2 \otimes \dots \otimes H_n
\eeq
with each $H_i\simeq \mC^K$. Choosing a usual basis $e_i$ in $\mC^K$ we can also write
\beq
	A=\sum_{j_1=1}^K \dots \sum_{j_n=1}^K A_{j_1\dots j_n}e_{j_1}\otimes\dots\otimes e_{j_n}\ .
\eeq
Each S-matrix is a linear operator acting on the tensor product of two of the spaces $H_i$,
\beq
	\hat S_{ij}\in End(H_i \otimes H_j)
\eeq
where we put a hat on $S$ to underline that it now has a matrix structure. This of course matches the notation from the part of this collection focussed on S-matrices \cite{Smcourse} where the S-matrix has four indices -- two for incoming particles and two for outgoing ones. A typical example that we will discuss in the next section is
\beq
	\hat S_{12}(p_1,p_2)=f(p_1,p_2) \hat I + g(p_1,p_2) \hat P_{12}
\eeq
where $\hat I$ is the identity operator and $\hat P$ is the permutation operator, i.e. $\hat P_{12}\; (e_a \otimes e_b)=e_b \otimes e_a$, while $f$ and $g$ are some explicit functions.
As another illustration, we can write the unitarity condition as
\beq
	\hat S_{12}(p_1,p_2) \hat S_{12}(p_2,p_1)=1
\eeq
and the highly important Yang-Baxter equation as
\beq
\label{ybes}
	\hat S_{12}(p_1,p_2) \hat S_{13}(p_1,p_3) \hat S_{23}(p_2,p_3)=
	\hat S_{23}(p_2,p_3) \hat S_{13}(p_1,p_3) \hat S_{12}(p_1,p_2) 
\eeq
and it is satisfied over the space $H_1 \otimes H_2 \otimes H_3$.

Now we are ready to write the Bethe equations for a theory with several particle types. They are similar to \eq{aba1part} but the product of S-matrices now acts on a state $A\in H_1 \otimes H_2 \otimes \dots \otimes H_n\ $,
\beq
\boxed{
\label{abagenA}
	e^{i p_k L}\hat S_{k,k+1}\hat S_{k,k+2}\dots \hat S_{k,n} \hat S_{k,1}\dots \hat S_{k,k-1}\;A=A, \ \ 
	\ k=1,\dots,n
	}
\eeq
The energy is again the sum of individual energies. 

Our main goal is to understand how to solve this equation.
Notice that the ordering in the product is also important since $S$ is a matrix\footnote{These equations again match well the picture of taking one particle around the circle. E.g. after we scatter the 1st particle through the 2nd one they both change flavours, then the 2nd is untouched and we scatter the 1st (with the flavour now different) through the third, etc. So once we take the first particle around the circle it can change the flavour and also all other particles can change the flavour. The result has to match the initial wavefunction which is exactly the statement in \eq{abagenA}.}.
So, the r.h.s. of \eq{abagenA} is an operator acting in the full space $ H_1 \otimes H_2 \otimes \dots \otimes H_n$. To solve the Bethe equations \eq{abagenA} we need to find its eigenvalues and eigenvectors. Fortunately, this is possible to do in a very efficient way using the fact that the S-matrix satisfies the Yang-Baxter equation. In the next several sections we will demonstrate how this works in concrete examples.

To finish the discussion, let us outline the derivation of the Bethe equations \eq{abagenA}.  First we will need to write the wavefunction in a more explicit form. This was already discussed to some extent in the article on S-matrices \cite{Smcourse} (for more details on this see \cite{Zamolodchikov} and the review \cite{Stijn}). To automatically take care of the (anti-)symmetrization for identical particles, let us introduce creation and annihilation operators $a_j,\;a^\dagger_j$, whose index $j$ labels the different particle types. Then we can describe the wavefunction as
\beq
	\Psi_{i_1\dots i_n}(x_1,\dots,x_n)=\bra{0}a_{i_1}(x_1)\dots a_{i_n}(x_n)\ket{\Psi(p_1,\dots,p_n)}
\eeq
where we use the product of annihilation operators to extract the part of the wavefunction corresponding to particle flavours $i_1,\dots,i_n$. The state $\ket{\Psi(p_1,\dots,p_n)}$ is defined as
\beqa
\label{PsiA}
	\ket{\Psi(\{p\})}&=&\int d^n y \sum_{\cP\in S_n}A^{\cP}_{j_1\dots j_n}(\{p\})
	\(\prod_{m=1}^n e^{ip_{\cP_m}y_m}\)
	\theta(y_1\ll\dots\ll y_n)\times
	\\ \nn
	&&
	a^\dagger_{j_1}(y_1)\dots a^\dagger_{j_n}(y_n)\ket{0}
\eeqa
where $\theta(x)$ is the Heaviside step function\footnote{I.e. $\theta(y_1\ll\dots\ll y_n)$ is equal to $1$ when $y_1\ll\dots\ll y_n$ and is equal to zero otherwise. Let us mention that this expression for the wavefunction is valid only in the regions when the particles are well separated, so it makes sense to consider the condition $y_1\ll\dots\ll y_n$. } and we assume summation over repeated indices. The coefficients $A$ are related to each other as
\beq
\label{APP}
	A^{\cP'}=\hat S_{\cP_i,\cP_{i+1}}\cdot A^\cP
\eeq
	if $\cP'$ is obtained from $\cP$ by permutation of elements $i,i+1$ corresponding to particles $\cP_i,\cP_{i+1}$. The flavour indices of the S-matrix in \eq{APP} are understood to be appropriately contracted with those of $A^\cP$. As any permutation can be written as a sequence of permutations that affect only two elements, any $A^\cP$ can be related to $A^{\cI}$ (with $\cI$ being the identity permutation) via a sequence of multiplication by the S-matrices. The Yang-Baxter equation satisfied by the S-matrix ensures this relation is unambiguous. 

As an example, for a state with two particles in a theory with only one particle type, we would get
\beq
	\Psi_{x_1 \ll x_2}=e^{ip_1x_1+ip_2x_2}+S(p_1,p_2)e^{ip_2x_1+ip_1x_2}
\eeq
\beq
	\Psi_{x_1 \gg x_2}=e^{ip_2x_1+ip_1x_2}+S(p_1,p_2)e^{ip_1x_1+ip_2x_2}
\eeq

{\textbf{Exercise:}} Derive these equations from \eq{PsiA}.

Let us now recall that we are considering the theory on a circle where the absolute ordering of particles is meaningless, so the ordering is only important up to cyclic permutations. Let us consider for example the wavefunction for $x_1\ll \dots \ll x_n$. If we define $y_1=x_1+L$ then since the separation between particles cannot be larger than $L$ we have $x_2\ll \dots\ll x_n\ll y_1$. For this ordering we would get from \eq{PsiA} a different expression for the wavefunction, but it should coincide with the first one as on a circle $x_1$ is indistinguishable from $x_1+L$. This leads to
\beq
	e^{-i p_1 L}A^\cI=\hat S_{1,2}\hat S_{1,3}\dots \hat S_{1,n}A^\cI
\eeq
and in general (notice that the ordering in the product is important since $S$ is a matrix)
\beq
	e^{-i p_k L}A^\cI=\hat S_{k,k+1}\hat S_{k,k+2}\dots \hat S_{k,n} \hat S_{k,1}\dots \hat S_{k,k-1}A^\cI
\eeq
These are precisely the equations \eq{abagenA} that were announced above, where $A$ is identified with $A^\cI$.


In conclusion, the crucial problem is to diagonalize the product of S-matrices in \eq{abagenA}. In the next section we will describe a general procedure for doing this based on the Yang-Baxter equation, and then we will see how it works for concrete examples. 
%

\section{Algebraic Bethe ansatz: building the transfer matrix}
\label{sec:aba}

The method we are going to use for solving the periodicity condition \eq{abagenA} goes under the name of the algebraic Bethe ansatz. In this section we will discuss its part which is common for all models -- the construction of the so-called transfer matrix -- and later we will specialize to concrete examples.

The key insight which allows to diagonalize the product of S-matrices in \eq{abagenA} is to introduce an {\textit{unphysical}} particle with momentum $p$ in an auxiliary space $H_a\simeq \mC^K$ and scatter it through all our particles. That is, we define the monodromy matrix
\beq
\label{Tdef1}
	\hat T_a(p)=\hat S_{a1}(p,p_1) \hat S_{a2}(p,p_2)\dots \hat S_{an}(p,p_n)
\eeq
which acts in $H_a \otimes H$ where $H=H_1 \otimes \dots \otimes H_n$ is our physical Hilbert space.
From the Yang-Baxter equation for the S-matrix it follows that the monodromy matrix satisfies a similar condition:
\beq
\label{STT1}
	\hat S_{ab}(p,p')\hat T_a(p)\hat T_b(p')=\hat T_b(p')\hat T_a(p)\hat S_{ab}(p,p')
\eeq
Then we define the transfer matrix by taking a trace over the auxiliary space
\beq
	\hat T(p)= \Tr_a \hat T_a(p),\ \ \ 
\eeq
and it is now an operator on the physical space only,
\beq
	\hat T(p) \in End(H_1 \otimes \dots \otimes H_n)\ .
\eeq
Remarkably, the transfer matrices for different values of $p$ commute,
\beq
	[\hat T(p),\hat T(p')]=0\ \ .
\eeq
This follows from the ``RTT relation'' \eq{STT1}\footnote{the name 'RTT relation' is due to the fact that in the literature usually one has the R-matrix in place of the S-matrix} and is the main point of the construction\footnote{Here is one way to explain the construction informally. We want to build a commuting set of operators on the Hilbert space $H$, and for this we uplift it to $H\otimes H_a \otimes H_b$, then on that space we have $\hat T_a$ and $\hat T_b$ which 'almost commute' -- up to multiplication by S-matrices as in \eq{STT1}. Then the operators obtained from $\hat T_a, \hat T_b$ by tracing over the auxiliary space will really commute with each other on the physical Hilbert space.}.
This commutativity means that they have a common set of eigenvectors. 

Moreover, $\hat T(p)$ is also related to the product of S-matrices that we want to diagonalize. To show this we need an extra property
\beq
	\hat S_{12}(p,p)=-\hat P_{12}
\eeq
which holds in many theories including all examples we consider below. Then
\beqa
	\hat T(p_1)&=&-\Tr_a \hat P_{a1}\hat S_{a2}(p_1,p_2)\dots \hat S_{an}(p_1,p_n)
		\\ \nn
	&=& -\Tr_a\hat S_{12}(p_1,p_2)\dots \hat S_{1n}(p_1,p_n)\hat P_{a1}
	\\ \nn
	&=&
	-\hat S_{12}(p_1,p_2)\dots \hat S_{1n}(p_1,p_n)
\eeqa
and the result is exactly the operator in the r.h.s. of the periodicity condition. We have used
\beq
	\hat P_{a1}\hat S_{ai}=\hat S_{1i}\hat P_{a1}, \ \ \ \Tr_a \hat P_{ab}=\hat I_b
\eeq
\beq
	\hat P_{ab}^2=1
\eeq
\beq
	\hat P_{ab}=\hat P_{ba}
\eeq
\beq
	\hat P_{a1}\hat P_{a2}=\hat P_{a2}\hat P_{12}=\hat P_{12}\hat P_{a1}
\eeq
Using cyclicity of the trace we can also show that for any $k$ the transfer matrix $\hat T(p_k)$ gives the operator that we want to diagonalize,
\beq
	\hat T(p_k)=-\hat S_{k,k+1}\hat S_{k,k+2}\dots \hat S_{k,n} \hat S_{k,1}\dots \hat S_{k,k-1}
\eeq
So our goal is to solve the eigenvalue problem for $\hat T(p)$,
\beq
	\hat T(p) A = \Lambda(p) A
\eeq
and then the periodicity condition \eq{abagenA} reduces to just an algebraic equation,
\beq
\label{eqperLam}
	e^{-ip_k L}=-\Lambda(p_k) \ \ !
\eeq
In the next sections we will attack the problem of diagonalizing the transfer matrix $\hat T(p)$ for several models.

\begin{figure}
\begin{center}
\includegraphics[scale=0.85]{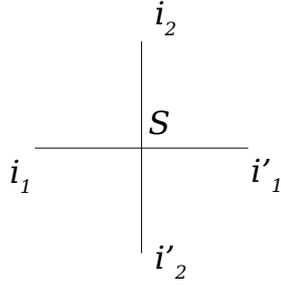}
\end{center}
\caption{\label{fig:sindices}\textbf{Graphical representation of the S-matrix.} The S-matrix $\hat S_{12}$ acting in $H_1\otimes H_2$ is shown as an intersection of two lines. Each of the two lines corresponds to one of the two spaces $H_1,H_2$. The four ends of the lines correspond to the four indices of the S-matrix.}
\end{figure}

\begin{figure}
\begin{center}
\includegraphics[scale=0.85]{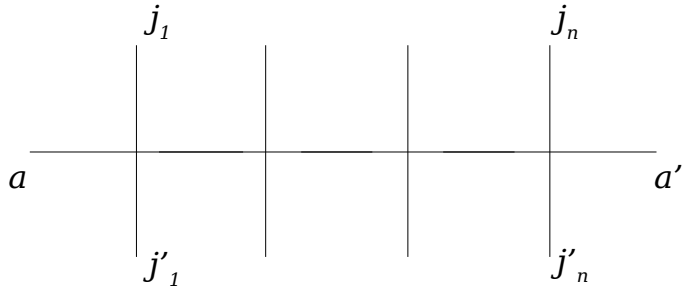}
\end{center}
\caption{\label{fig:transfmat}\textbf{Graphical representation of the monodromy matrix $\hat T_a$.} The monodromy matrix is a product of several S-matrices. The horizontal line corresponds to the auxiliary space $H_a$, while vertical lines are associated with the physical spaces $H_1,\dots,H_n$.}
\end{figure}

\begin{figure}
\begin{center}
\includegraphics[scale=0.8]{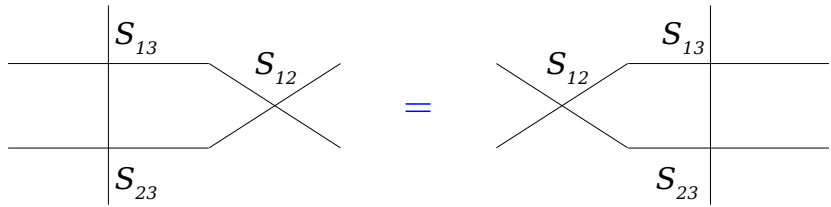}
\end{center}
\caption{\label{fig:ybe}\textbf{Graphical representation of the Yang-Baxter equation \eq{ybes}.} The equation means that we can move the vertical line across the intersection point of the two other lines.}
\end{figure}

\begin{figure}
\begin{center}
\includegraphics[scale=0.7]{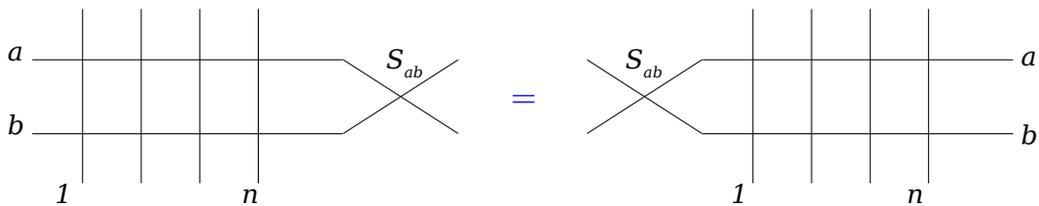}
\end{center}
\caption{\label{fig:rtt}\textbf{The RTT relation \eq{STT1}.} The two horizontal lines correspond to auxiliary spaces $H_a,H_b$, while the vertical lines correspond to physical spaces $H_1,\dots,H_n$ (labels on the picture next to the lines show which spaces the lines are associated with). Due to the Yang-Baxter equation we can move all the vertical lines one by one to the other side of the intersection, leading to the RTT relation \eq{STT1}.}
\end{figure}

Let us mention that a pictorial representation is often used for the transfer matrix and the S-matrix, as already discussed in part in other chapters of this collection \cite{Yangiancourse,Smcourse}. To understand how it works, let us write the S-matrix $\hat S_{12}$ in index notation. As it acts on the tensor product of two spaces $H_1\otimes H_2$, its index structure is $S^{i_1i_2}_{i'_1i'_2}$ where the first upper index and the first lower index correspond to the $H_1$ space and the second pair of indices corresponds to the $H_2$ space. As shown on Fig. \ref{fig:sindices}, we can represent this structure a pair of intersecting lines, with the ends of one line corresponding to the $H_1$ space and the ends of the other line to the $H_2$ space. The convenience of this notation is that contraction of indices in any expression with a product of operators such as \eq{Tdef1} is simply represented as joining the corresponding lines. For example the monodromy matrix can be depicted as shown on Fig. \ref{fig:transfmat}. The Yang-Baxter equation in graphical form is shown on Fig. \ref{fig:ybe}. Notice that using the pictorial representation one can easily prove the RTT relation using the Yang-Baxter equation, as shown on Fig. \ref{fig:rtt}.

\section{Algebraic Bethe ansatz: solving the SU(2) chiral Gross-Neveu model}
\label{sec:su2gn}

We will now specialize to a concrete example: the chiral $SU(2)$ Gross-Neveu model (in section \ref{sec:su3} we will study the more complicated $SU(3)$ case). This theory has already been discussed and introduced in the part of this collection about S-matrices \cite{Smcourse}. We will parameterize the particles' energy and momentum\footnote{here we choose an unconventional prefactor in front of $u$ for a better match with the usual spin chain notation} in terms of the rapidity $u$,
\beq
\label{pEsu2}
	E=m \cosh \frac{\pi u}{2},\ \ \ p=m \sinh \frac{\pi u}{2}
\eeq
In the $SU(2)$ chiral Gross-Neveu model there are effectively two massive particles, so in terms of the previous notation $K=2$ and the S-matrix acts in $\mC^2 \otimes \mC^2$. Explicitly, the S-matrix reads 
\beq
\label{SandRsu2}
	\hat S_{12}(p_1,p_2)=S^{ff}(u_1-u_2)\hat R_{12}^{-1}(u_1-u_2)
\eeq
\beq
	S^{ff}(u)=-\frac{\Gamma(1-\frac{u}{4i})\Gamma(\frac{1}{2}+\frac{u}{4i})}
	{\Gamma(1+\frac{u}{4i})\Gamma(\frac{1}{2}-\frac{u}{4i})}
\eeq
where
\beq
	\hat R_{12}(u)=\ofrac{u+2i}\(u \hat I + 2i \hat P_{12}\)\ .
\eeq
This is the usual R-matrix of the rational type, satisfying the Yang-Baxter equation
\beq
	\hat R_{12}(u-v) \hat  R_{1a}(u)\hat  R_{2a}(v)=\hat R_{2a}(v)\hat R_{1a}(u)\hat R_{12}(u-v)
\eeq
as well as
\beq
	\hat R_{12}(u) \hat R_{12}(-u)=1\ .
\eeq
As we discussed we need to diagonalize the transfer matrix built as a product of these S-matrices with a trace over the auxiliary space. For that the important thing is the matrix structure, so let us drop for some time the prefactor $S^{ff}(u)$, and then
\beq
	\hat T(u)=\Tr_a \prod_{i=1}^{N_f}\hat R_{ai}^{-1}(u-u_i)
	=\Tr_a \prod_{i=1}^{N_f}\hat R_{ai}(u_i-u)
\eeq
where $N_f$ is the number of particles on the circle.

For some time now we will concentrate on the problem of diagonalizing this transfer matrix. Later we will return to the Gross-Neveu model and assembling all the ingredients we will write equations for its spectrum.

As we saw, the operators $\hat T(u)$ commute for different values of $u$. As the transfer matrix in our case is a polynomial, its operator coefficients in front of the powers of $u$ all commute with each other. In particular, if we set all $u_i$ to zero, a particular combination of these operators gives the famous XXX spin chain Hamiltonian  -- a system of $N_f$ interacting spins $s=1/2$. It is defined as
\beq
\label{HX}
	\hat H=-\frac{1}{4}\sum_{i=1}^{N_f}\(\vec{\sigma}_i\vec{\sigma}_{i+1}-1\)
\eeq
where $\sigma_i$ are the Pauli matrices acting on the $i$-th site and we identify $i=N_f+1$ and $i=1$. 
The same operator is expressed via the transfer matrix as
\beq
\label{HXXXT}
	\hat H=\left. i\frac{d}{du}\log \hat T(u)\right|_{u=0}\ .
\eeq
The Hamiltonian can be also written in terms of permutation operators,
\beq
	\hat H=\frac{1}{2}\sum_{k=1}^{N_f}\(1-\hat P_{k,k+1}\)\ .
\eeq
The importance of this Hamiltonian was first recognized in condensed matter applications where it serves as a model for a ferromagnetic material. In the AdS/CFT context the XXX Hamiltonian is also directly relevant as it describes the leading order anomalous dimensions for operators in a simple subsector of the $\cN=4$ supersymmetric Yang-Mills theory (see the review \cite{Minahan:2010js}).

If we keep $u_i$ nonzero, the relation \eq{HXXXT} gives the Hamiltonian for the spin chain for which $u_i$ are called the inhomogenieties. The fact that all coefficients of $\hat T(u)$ commute with the Hamiltonian, i.e. represent a large number of conservation laws, is a strong sign for the integrability of this model and ultimatley leads to its solution.

To construct the eigenstates of $\hat T(u)$ we will use operators originating from the monodromy matrix $\hat T_a(u)$. We can write the monodromy matrix explicitly as a $2\times 2$ matrix in auxiliary space,
\beq
	\hat T_a(u)=
	\begin{pmatrix}
	\hat A(u) & \hat B(u) \\
	\hat C(u) & \hat D(u) \\
	\end{pmatrix}
\eeq
where the entries act on the physical space,
\beq
	\hat A(u), \hat B(u), \hat C(u), \hat D(u) \in End(H_1 \otimes \dots \otimes H_n).
\eeq
In this notation we have
\beq
	\hat T(u)=\hat A(u)+\hat D(u)\ .
\eeq
The entries of $\hat T(u)$ satisfy important commutation relations following from the identity \eq{STT1} which takes the form\footnote{the R-matrix has argument $v-u$ as $\hat T_a$ is built from $\hat R^{-1}$}
\beq
	\hat R_{12}(v-u)\hat T_1(u)\hat T_2(v)=\hat T_2(v)\hat T_1(u)\hat R_{12}(v-u)\ .
\eeq
In particular,
\beq
	[\hat B(u),\hat B(v)]=0
\eeq
\beq
\label{ABsu2}
	\hat A(v)\hat B(w)=\frac{v-w+2i}{v-w}\hat B(w)\hat A(v)-\frac{2i}{v-w}\hat B(v)\hat A(w)
\eeq
\beq
\label{DBsu2}
	\hat D(v)\hat B(w)=\frac{w-v+2i}{w-v}\hat B(w)\hat D(v)+\frac{2i}{v-w}\hat B(v)\hat D(w)
\eeq
\textbf{Exercise:} Derive these relations.

Let us introduce the vacuum state $\ket{0}$, in which all spins are up\footnote{Sometimes this state is called `pseudovacuum' rather than vacuum, since e.g. the ground state in this model is actually degenerate, for instance the state with all spins down has the same energy.}:
\beq
	\ket{0}=\ket{\uparrow\uparrow\dots \uparrow}=
	\begin{pmatrix}1\\0\end{pmatrix}\otimes\dots\begin{pmatrix}1\\0\end{pmatrix}
\eeq
Using the explicit form of the R-matrix we find that the vacuum is an eigenstate of $\hat T$ and
\beq
	\hat A(u)\ket{0}=\ket{0},\ \ 
	\hat D(u)\ket{0}=\prod_j \frac{u_j-u}{u_j-u+2i}\ket{0},\ \ 
	\hat C(u)\ket{0}=0\ .
\eeq
The idea is to view $\hat B$ as a creation operator and build the transfer matrix eigenstates as
\beq
	\ket{w_1,\dots,w_{N_a}}=\hat B(w_1)\hat B(w_2)\dots \hat B(w_{N_a})\ket{0}
\eeq
where $w_1,w_2,\dots$ are parameters known as Bethe roots. Let's see how $\hat T=\hat A+\hat D$ acts on this state. If the second term in the r.h.s. of \eq{ABsu2} was absent, we could just commute $\hat A(u)$ through all $\hat B$'s until it hits the vacuum which is its eigenstate. Similarly if there was no second term in the r.h.s. of \eq{DBsu2} we could commute $\hat D(u)$ through all $\hat B$'s and then again arrive at the vacuum. However due to the presence of these extra terms in \eq{ABsu2}, \eq{DBsu2} we will find extra \textit{unwanted} contributions, and the full result is
\beqa
	\hat A(u)\ket{w_1,\dots,w_{N_a}}
	&=&\prod_j\frac{u-w_j+2i}{u-w_j}\hat B(w_1)\dots \hat B(w_{N_a})\ket{0}\\ \nn
	&+&\sum_j M_j \hat B(u)\hat B(w_1)\dots \hat B(w_{j-1})\hat B(w_{j+1})\dots\hat B(w_{N_a})\ket{0}
\eeqa
\beqa
	\hat D(u)\ket{w_1,\dots,w_{N_a}}
	&=&\prod_j\frac{w_j-u+2i}{w_j-u}\prod_k\frac{u_k-u}{u_k-u+2i}
	\hat B(w_1)\dots \hat B(w_{N_a})\ket{0}\\ \nn
	&+&\sum_j \tilde M_j \hat B(u) \hat B(w_1)\dots \hat B(w_{j-1})\hat B(w_{j+1})\dots\hat B(w_{N_a})\ket{0}
\eeqa
The unwanted terms in these two equations are those that include $M_j, \tilde M_j$. Explicitly, by $M_j$ we denote the coefficient of the term where $\hat A$ was commuted with $\hat B$ using the second term in the commutation relation \eq{ABsu2}, i.e. the one which exchanges the arguments (and similarly for $\tilde M_j$). For instance, it is easy to compute
\beq
	M_1=\frac{-2i}{u-w_1}\prod_{k\neq 1}\frac{w_1-w_k+2i}{w_1-w_k}
\eeq
\beq
	\tilde M_1=\frac{2i}{u-w_1}
	\prod_{m}\frac{u_m-w_1}{u_m-w_1+2i}\prod_{k\neq 1}\frac{w_k-w_1+2i}{w_k-w_1}
\eeq
And since all $\hat B$'s commute, $M_j,\tilde M_j$ are trivial generalizations of these expressions,
\beq
	M_j=\frac{-2i}{u-w_j}\prod_{k\neq j}\frac{w_j-w_k+2i}{w_j-w_k}
\eeq
\beq
	\tilde M_j=\frac{2i}{u-w_j}
	\prod_{m}\frac{u_m-w_j}{u_m-w_j+2i}\prod_{k\neq j}\frac{w_k-w_j+2i}{w_k-w_j}
\eeq
We see that for any $j$ we can cancel the unwanted terms against each other! This will happen if
\beq
	\prod_{k\neq j}\frac{w_j-w_k+2i}{w_j-w_k}=
	\prod_{m}\frac{u_m-w_j}{u_m-w_j+2i}\prod_{k\neq j}\frac{w_k-w_j+2i}{w_k-w_j}
\eeq
It's convenient to relabel the Bethe roots as $\tilde w_k=w_k+i$, then dropping the tilde we get
\beq
\label{abasu21}
	\prod_{m}\frac{w_j-u_m+i}{w_j-u_m-i}=\prod_{k\neq j}\frac{w_j-w_k+2i}{w_j-w_k-2i}\ .
\eeq
The equations \eq{abasu21} are known as Bethe equations for the XXX chain, and they are one of the key results of this section. 
The eigenvalue of the transfer matrix then reads
\beq
\label{lamsu21}
	\Lambda_{SU(2)}(u)=\prod_m\frac{u-w_m+i}{u-w_m-i}+
	\prod_k\frac{u-u_k}{u-u_k-2i}\prod_m\frac{u-w_m-3i}{u-w_m-i}
\eeq
This is the main outcome of our discussion. In particular, we can extract from it the eigenvalue of the XXX Hamiltonian
\eq{HX}. To do this we use the relation \eq{HXXXT} which links the Hamiltonian to the transfer matrix in which all $u_i$ should be set to zero. We find the simple result\footnote{In the spin chain literature the Bethe roots $w_k$ and the inhomogenieties $u_m$ are usually rescaled by a factor of two compared to our notation so that the Bethe equations would read $
	\prod_{m}\frac{w_j-u_m+i/2}{w_j-u_m-i/2}=\prod_{k\neq j}\frac{w_j-w_k+i}{w_j-w_k-i}
$
and the energy of the XXX Hamiltonian would be $
	E=\frac{1}{2}\sum_j\frac{1}{w_j^2+1/4}
$
}
\beq
\label{Exxxsu2}
	E=2\sum_j\frac{1}{w_j^2+1}\ .
\eeq
Also, it is important that the XXX spin chain Hamiltonian has an $SU(2)$ symmetry, which will be discussed in more detail in section \ref{sec:states}, together with its implications for the structure of the eigenstates and eigenvalues.

Let us finally underline that the Bethe equations we have just obtained give the spectrum of the transfer matrix and of the spin chain Hamiltonian \textit{exactly} at any length of the chain, i.e. any $N_f$. This is in contrast with the Bethe ansatz for the 2d field theory we started with, which captures only powerlike and not exponential corrections in the volume $L$.

One can ask whether these equations provide ${\textit{all}}$ eigenstates and eigenvalues of the transfer matrix, i.e.  whether the algebraic Bethe ansatz solution is complete. While the answer is certainly expected to be positive, a fully rigorous proof has not been found so far (the proofs which are available rely on some conjectures, see \cite{Faddeev} for an initial discussion and also \cite{Hao:2013jqa} as well as references therein for a more recent summary). A related issue is that the Bethe ansatz could have some singular solutions which do not correspond to eigenstates. This issue as well as the question of completeness become more tractable if one introduces twisted boundary conditions for the spinj chain (see e.g. the recent discussion in \cite{Nepomechie:2014hma} and references therein). It is also expected that it is sufficient to consider only solutions where the  Bethe roots are pairwise distinct in order to get all eigenvalues of the Hamiltonian.


Let us mention that there is a shortcut to the Bethe equations for our transfer matrix. Suppose we forget about the unwanted terms in the commutation relations, then we would still arrive at the same expression for the eigenvalue \eq{lamsu21}. This eigenvalue however appears to have  poles when $u=w_j$. The poles cannot be really there as the transfer matrix is not singular at these points\footnote{it's only singular at $u=u_j+2i$ which corresponds to the pole in our R-matrix}. Demanding that the residue of the poles vanish we obtain equations on the roots $w_k$ -- which are nothing but the Bethe ansatz equations \eq{abasu21} ! This is not a rigorous derivation of the Bethe equations, but this trick is very useful. We will apply it in section
\ref{sec:su3} for the $SU(3)$ case. 

Notice that if we had only the XXX Hamiltonian it would be very hard to guess the transfer matrix and the algebraic Bethe ansatz procedure! Historically the XXX chain was solved first by another method which we will discuss in the next section.

\subsection{Coordinate Bethe ansatz for the XXX Hamiltonian}
\label{sec:cba}

The exact solution of the XXX chain was originally obtained by a more intuitive method known as the coordinate Bethe ansatz \cite{Bethe:1931hc} (see e.g. \cite{Faddeev} for a review and more details on this model). Let us forget about the transfer matrix and consider just the XXX chain Hamiltonian,\footnote{In this subsection as well as in section 5 below we denote the length of the chain as $L$ (to simplify notation) rather than as $N_f$ which was used in the discussion above.}
\beq
	\hat H=\frac{1}{2}\sum_{k=1}^{L}\(1-\hat P_{k,k+1}\)
\eeq
where as usual we identify the $(L+1)$-th and the 1st sites (we consider the case when there are no inhomogenieties, $u_i=0$). The method involves making a clever guess ({\textit{ansatz}) for the explicit form of the eigenstates. We start with the ground state in which all spins are up,
\beq
	\hat H \ket{\upa\dots\upa}=0\ .
\eeq
Let us look for the first excited state as a combination of terms
\beq
	\ket{n}=\ket{\upa\upa\dots\upa\dna\upa\dots\upa}
\eeq
where one spin at the $n$th site is flipped. Writing
\beq
	\ket{\Psi}=\sum_n e^{ipn}\ket{n}\ ,
\eeq
and noting also the periodic boundary conditions $n\sim n+L$ we find that it's an eigenstate if
\beq
\label{cbaper1}
	e^{ipL}=1 \ .
\eeq
The corresponding energy is
\beq
\label{E1magn}
	E(w)=\frac{2}{w^2+1}\ ,
\eeq
where we parameterize the momentum as
\beq
	e^{ip}=\frac{w+i}{w-i}\ .
\eeq
So it is natural to understand $\ket{\psi}$ as a 1-particle state, and the momentum of the particle is quantized according to \eq{cbaper1}. Also, notice that the energy \eq{E1magn} matches the general formula \eq{Exxxsu2} for the case with only one root $w_j$.

Let us further write a two-particle state as
\beq
	\ket{\Psi}=\sum_{1\leq n<m\leq L}\psi(n,m)\ket{n,m}\ ,
\eeq
where $\ket{n,m}$ is the state with $n$th and $m$th spins flipped. We make an ansatz for the wavefunction $\psi$ as
\beq
	\psi(n,m)=e^{ip_1n+ip_2m}+S(p_1,p_2)e^{ip_1m+ip_2n}\ ,
\eeq
where the coefficient $S$ is to be understood as a phase acquired by the wavefunction when the two particles scatter through each other. We find that this will be an eigenstate for
\beq
	S(p_1,p_2)=\frac{w_1-w_2+2i}{w_1-w_2-2i}\ ,
\eeq
with the eigenvalue $E(u_1)+E(u_2)$.
What is truly remarkable is that this construction still works for more than two particles. E.g. for three excitations, denoting
\beq
	\ket{p_1,p_2,p_3}=\sum_{1\leq n_1<n_2<n_3\leq L}e^{ip_1n_1+ip_2n_2+ip_3n_3}\ket{n_1,n_2,n_3}\ ,
\eeq
we can write the wavefunction as
\beqa
	\ket{\psi}&=&\ket{p_1,p_2,p_3}+S_{12}\ket{p_2,p_1,p_3}
	+S_{23}\ket{p_1,p_3,p_2}+S_{13}S_{12}\ket{p_2,p_3,p_1}\\ \nn
	&+&S_{13}S_{23}\ket{p_3,p_1,p_2}
	+S_{12}S_{13}S_{23}\ket{p_3,p_2,p_1}
\eeqa
(with $S_{ij}=S(p_i,p_j)$) and it is still an eigenstate provided the Bethe equations
\beq
\label{abasu2std}
	\(\frac{w_j+i}{w_j-i}\)^L=\prod_{k\neq j}\frac{w_j-w_k+2i}{w_j-w_k-2i}
\eeq
are satisfied\footnote{These equations are of course obtained from \eq{abasu21} by setting all $u_i$ to zero.}. Notice that the wavefunction is built using only the same two-particle scattering S-matrix, so in this sense multiparticle scattering is reduced to only $2\to 2$ interactions. This is in full analogy with the factorization of scattering in 2d integrable QFTs. This spin chain wavefunction is in fact one of the inspirations for writing the QFT wavefunction in the form that we did before.



\subsection{Bethe equations for the $SU(2)$ chiral Gross-Neveu model}

Let us get back to the $SU(2)$ Gross-Neveu model. Now we are ready to write the full set of Bethe equations for its spectrum. To do this we should plug the explicit expression \eq{lamsu21} for the transfer matrix eigenvalue $\Lambda$ into the periodicity condition \eq{eqperLam}. Adding the scalar factor $S^{ff}(u)$ that we dropped before, we get
\beq
	e^{ip_jL}\prod_mS^{ff}(u_j-u_m)\prod_k\frac{u_j-w_k+i}{u_j-w_k-i}=-1\ .
\eeq
These equations should be supplemented by the Bethe equations \eq{abasu21} which fix the parameters $w_j$,
\beq
	\prod_{m}\frac{w_j-u_m+i}{w_j-u_m-i}=\prod_{k\neq j}\frac{w_j-w_k+2i}{w_j-w_k-2i}\ .
\eeq
Then the energy is given by
\beq
	E=\sum_j m \cosh \frac{\pi u_j}{2}\ .
\eeq
We expect this to be the {\textit{exact}} result for the energy, up to corrections that are exponentially small in $L$. This concludes our solution for the spectrum of the Gross-Neveu model at large $L$. 

In the next section we will discuss the XXX chain in some more detail, and then in section \ref{sec:su3} derive a generalization of these equations to the $SU(3)$ Gross-Neveu model.

\section{Exploring the XXX spin chain}

The $SU(2)$ XXX spin chain which we already encountered in the previous section is a very important and widely used model, and deserves a deeper look. In this section we will discuss several of its features in more detail. We will consider the case without any inhomogeneities for clarity.

\subsection{The Bethe states in-depth}
\label{sec:states}

In the previous section we saw that the XXX spin chain Hamiltonian
\beq
	\hat H=\frac{1}{2}\sum_{k=1}^{L}\(1-\hat P_{k,k+1}\)
\eeq
can be diagonalized via algebraic Bethe ansatz and its eigenstates are built as\footnote{To simplify notation compared to the discussion of the Gross-Neveu model above, in this section we denote the length of the chain as $L$ and the number of excitations as $M$.}
\beq
\label{bbvac}
	\ket{\Psi}=\hat B(w_1)\hat B(w_2)\dots \hat B(w_M)\ket{\upa\upa\dots\upa}\ ,
\eeq
where the Bethe roots are determined by
\beq
	\(\frac{w_j+i}{w_j-i}\)^L=\prod_{k\neq j}^M\frac{w_j-w_k+2i}{w_j-w_k-2i}\ .
\eeq
Importantly, this Hamiltonian commutes with the operators $\hat S_x,\hat S_y,\hat S_z$ giving the total spin of the system, which are defined as a sum of the individual spins, i.e.
\beq
\label{commSH}
	[\hat H, \hat S_x]=[\hat H, \hat S_y]=[\hat H, \hat S_z]=0,\ 
\eeq
\beq
	\hat S_\alpha=\sum_{i=1}^L \hat S_{\alpha}^{(i)},\ \ \ \alpha=x,y,z\ \ .
\eeq
These operators $\hat S_x,\hat S_y,\hat S_z$ are the generators of the global $SU(2)$ symmetry algebra under which the Hamiltonian is thus invariant. First, this means that we can choose eigenstates of $\hat H$ to be eigenstates for $\hat S_z$ as well. Also, if we have an eigenstate we can generate more eigenstates (with the same energy) by acting on it repeatedly with the spin-lowering or raising operators $\hat S_{\pm}=\hat S_x\pm i \hat S_y$.

Since there is an $SU(2)$ algebra acting at each site of the chain, the whole $2^L$-dimensional Hilbert space is a tensor product of $L$ copies of the fundamental representation of $SU(2)$. It can be thus decomposed into a direct sum of irreducible representations (irreps) $V_\alpha$ of the global $SU(2)$ symmetry,
\beq
	\mC^2\otimes\dots\otimes\mC^2=\oplus_{\alpha}V_\alpha \ .
\eeq
Due to the relations \eq{commSH} above, each $V_\alpha$ is an invariant subspace for the Hamiltonian and all states there have the same energy. For instance, when $L=2$ we have a system of two spin-1/2 particles, so the Hilbert space decomposes into a spin-0 and a spin-1 representation. The full space is spanned by the three states
\beq
\label{spin1s}
	\ket{\upa\upa},\ \ket{\dna\dna},\ \ofrac{\sqrt{2}}\(\ket{\upa\dna}+\ket{\dna\upa}\)
\eeq
which form the spin-1 irrep of the global $SU(2)$, together with the state
\beq
\label{spin0s}
	\ofrac{\sqrt{2}}\(\ket{\upa\dna}-\ket{\dna\upa}\)
\eeq
in the singlet (spin-0) representation. The three states \eq{spin1s} all have zero energy like the ground state (which is one of them), while the singlet state has the energy $E=2$.

In each of the representations $V_\alpha$ there is a highest weight state from which all other states are obtained by acting with the lowering operator $\hat S_-$. These other states are known as descendants. In our example above with $L=2$, the states $\ket{\upa\upa}$ and $\ofrac{\sqrt{2}}\(\ket{\upa\dna}-\ket{\dna\upa}\)$ are highest weight. In fact, the highest weight states are always precisely the Bethe states of the algebraic Bethe ansatz \eq{bbvac}.\footnote{And conversely, the states built as \eq{bbvac} with finite $w_k$ are highest weight states.}

{\textbf{Exercise.}} What are the solutions of Bethe equations corresponding to the highest weight states for $L=2$ ?

There is also a very handy descripton for the descendants, as the lowering operator is expressed through the operator $\hat B(w)$ in the limit $w\to\infty$,
\beq
\label{Binf}
	\hat B(w)={\rm{const}} \times w^{L-1} \hat S_-+\dots,\ \ w\to\infty\ .
\eeq
It's easy to see that one can add roots $w_k=\infty$ to any solution of the Bethe ansatz and the Bethe equations will still be satisfied. Thus for finite roots the Bethe state \eq{bbvac} is a highest weight state, and if some roots are at infinity it is a descendant. Also, due to 
\beq
	[\hat S_z,\hat B(u)]=-\hat B(u)
\eeq
the Bethe state \eq{bbvac} is always an eigenstate of $\hat S_z$, with eigenvalue $L/2-M$, and we can think of this state as having $M$ spins flipped from $\ket{\upa}$ to $\ket{\dna}$.


As we said above the Bethe ansatz is expected to give the complete spectrum for this model. Also, all highest weight eigenstates can be constructed as Bethe states \eq{bbvac}. In fact to get all highest weight states it's enough to consider only $M\leq L/2$ in the Bethe equations (this is clear from the fact that the eigenvalue of $\hat S_z$ for a Bethe state is $L/2-M$). 

Let us also mention that the algebraic and coordinate Bethe ansatz solutions both rely on the existence of a simple reference eigenstate ($\ket{0}$ in our case), on top of which the other states are constructed. In some integrable models like the anisotropic XYZ chain such a simple reference state is not known in the generic situation, and one has to use other methods to solve them (see \cite{Cao:2013gug} for some recent discussion).

Another curious fact is that the Bethe equations have a {\textit{potential}}. More precisely there exists a function $F(\{w_j\})$ such that the Bethe equations are obtained from its derivative. To see this let us take the logarithm of the Bethe equations, finding
\beq
\label{baelog}
L \log\frac{w_j+i}{w_j-i}=\sum_{k=1,k\neq j}^J\frac{w_j-w_k+2i}{w_j-w_k-2i}+2\pi i n_j\ ,
\eeq
where the integers $n_j\in\mZ$ are known as mode numbers\footnote{Sometimes not $n$ but $-n$ is called the mode number in the literatureю}. Then we can write these equations for \textit{all} values of $k$ as a derivative of a single function,\footnote{One could alternatively include in $F$ an extra term $-2\pi i\sum_k  n_k w_k$ with some specific $n_k$, then the Bethe equations would read just $\d F/\d w_k=0$ but $F$ would explicitly depend on the specific choice of the mode numbers $n_k$ which are in general different for different states.}
\beq
	\frac{\d F}{\d w_k}=2\pi i n_k,\ \ n_k \in\mZ,\ \ \ k=1,\dots,M
\eeq
where
\beqa
	F&=&L\sum_{j=1}^M \[(w_j+i)\log(w_j+i)- (w_j-i)\log(w_j-i)\]\\ \nn
	&&+\sum_{k<j}^M\[(w_k-w_j-2i)\log(w_k-w_j-2i)-(w_k-w_j+2i)\log(w_k-w_j+2i)\]\ .
\eeqa
The function $F$ is known as the Yang-Yang function (its analog first appeared in \cite{Yang}) and can be generalized to almost any other quantum integrable model. In particular it plays an important role in the relation between ${\mathcal N=2}$ supersymmetric gauge theories in four dimensions and integrable systems \cite{Nekrasov}. Some of its properties are further discussed in the lecture course of this collection devoted to Thermodynamic Bethe Ansatz \cite{TBAcourse}.

Let us finally mention for completeness the celebrated Gaudin formula for the norm of the Bethe states \cite{Gaudin,Korepin}. It can be derived almost solely from the commutation relations between elements of the transfer matrix. The result is written as\footnote{This formula is valid if the set of Bethe roots is invariant under complex conjugation, i.e. $\{w_j\}=\{w_j\}^*$}
\beq
	\langle\Psi|\Psi\rangle=
	(2i)^M\prod_j \frac{w_j}{w_j+2i}\prod_{j\neq k}\(1+\frac{4}{(w_j-w_k)^2}\)
	\det_{m,n}\frac{\d^2 F}{\d w_m\d w_n}
\eeq
where the key part is the determinant involving the Yang-Yang function.
There is also a generalization of this formula, again in determinant form, for a scalar product of two Bethe states in one of which the $u_k$ are ``off-shell'', i.e. do not necessarilly satisfy the Bethe equations. Finding a compact extension of that formula to a higher rank (e.g. $SU(3)$) chain is a famous and longstanding open problem.

\subsection{Spectral curve and finite-gap equations}
\label{sec:curve}

Let us now discuss some interesting features which emerge in the \textit{classical} limit of the XXX chain -- namely the limit when the number of excitations and the length are very large while their ratio is finite. We will see that in this limit the Bethe equations reduce to a set of discontinuity conditions known as the finite-gap equations which are ubiquitous in classical (rather than quantum) integrable systems. These equations also define a Riemann surface known as the classical spectral curve, which encodes the conserved charges of the system. For a particular simple example we will show how to solve these equations and compute the energy using only various  analyticity constraints. For a more detailed discussion we refer the reader to e.g. \cite{KMMZ}\footnote{due to our notation there are various factor of 2 differences with \cite{KMMZ}}. Our discussion here complements the description of classical integrable systems in another part of the present collection \cite{Cintcourse}.

It is convenient to use the Bethe equations in logarithmic form \eq{baelog}, i.e.
\beq
L \log\frac{w_j+i}{w_j-i}=\sum_{k=1,k\neq j}^M\frac{w_j-w_k+2i}{w_j-w_k-2i}+2\pi i n_j
\eeq
In our limit $L$ and $M$ are very large.  
The roots will scale as $w_j\sim L$, and let us use the rescaled roots defined as
\beq
	w_j=L x_j,\ \ \ x_j\sim 1
\eeq
then we have
\beq
\label{baex}
	\frac{1}{x_j}=\frac{2}{L}\sum_{k\neq j}\ofrac{x_j-x_k}+\pi n_j\ .
\eeq
The number of roots is very large and they will get close to each other, so instead of a discrete set one can describe them as a continous distribution. The roots will form several cuts in the complex plane. Let us introduce the density of the roots
\beq
	\rho(x)=\ofrac{L}\sum_j\delta(x-x_j)
\eeq
and the resolvent
\beq
\label{defG}
	G(x)=\ofrac{L}\sum_j\ofrac{x-x_j}=\int_C\frac{d\xi\rho(\xi)}{x-\xi}\ ,
\eeq
where $C$ is the union of cuts $C_i$ on which the roots condense, $C=\cup C_i$. Introducing the filling fraction
\beq
	\alpha=M/L
\eeq
we have the normalization condition
\beq
	\int_C\rho(\xi)d\xi=\alpha\ ,
\eeq
so that
\beq
\label{Gas}
	G(x)=\frac{\alpha}{x}+\dots,\ \ \ x\to\infty\ .
\eeq
In our new variables the Bethe ansatz equations \eq{baex} can be written as
\beq
\label{baeG}
 	G(x+i0)+G(x-i0)=2\dint\frac{d\xi \rho(\xi)}{x-\xi}=\ofrac{x}-\pi n_j,\ \ \ \ x\in  C_j
\eeq
where the dashed integral sign means that we should take the integral's principal value (there is a singularity at $x=\xi$). These relations are known as finite-gap equations\footnote{the word 'finite' refers to the fact that we consider the case where the number of cuts $C_i$ is finite}. The resolvent has branch cuts formed by the Bethe roots and is thus a multivalued function whose Riemann surface is known as the classical spectral curve of the model. We derived it from a limiting case of the Bethe ansatz for the quantum XXX chain, but similar curves arise in various other situations, e.g. in classical integrable systems and in matrix models.

Let us also further restrict the solution by imposing an extra condition
\beq
	\prod_j\frac{w_j+i}{w_j-i}=1
\eeq
where each term in the product is nothing but $e^{ip_j}$ with $p_j$ being the momentum of a single excitation. This condition (known as the ``zero-momentum'' requirement) in fact means that the Bethe state is invariant under cyclic shifts of the sites. In our limit we can write it as
\beq
	P\equiv\ofrac{L}\sum_j \ofrac{x_j}=\pi m,\ \ m\in \mathbb{Z}\ ,
\eeq
or 
\beq
	\int d\xi \frac{\rho(\xi)}{\xi}=\pi m\ .
\eeq
Since at small $x$ we have
\beq
\label{Gsmallx}
	G(x)=-\int_C\frac{d\xi\rho(\xi)}{\xi}-x\int_C\frac{d\xi\rho(\xi)}{\xi^2}+\cO(x^2)\ ,
\eeq
we can equivalently write
\beq
\label{Gat0}
	G(0)=-\pi m\ .
\eeq

Let us now consider in detail an example when there is only one cut, whose mode number we denote as $n$.  We will fix the resolvent purely from analyticity constraints. From the definition of the resolvent we see that it is an analytic function with $1/x$ asymptotics at $x\to\infty$ and the only singularity being the branch cut formed by the Bethe roots. The discontinuity on the cut is
\beq
	G(x+i0)-G(x-i0)=-2\pi i\rho(x),\ \ \ x\in C
\eeq
Furthermore from \eq{baeG} we see that the values of the resolvent above and below the cut sum up to a meromorphic function, thus the cut is of the square root type. Combining all this we can write
\beq
\label{GfgQ}
	G(x)=f(x)+g(x)\sqrt{Q(x)}
\eeq
where $f,g$ are meromorphic functions and $Q$ is a polynomial. Since there are only two branch points, $Q$ has degree 2. From \eq{baeG} we get
\beq
	f(x)=\ofrac{2}\(\ofrac{x}-\pi n\)\ .
\eeq
With this $f(x)$ the resolvent $G(x)$ has an apparent singularity at $x=0$ which can only be  compensated by $g(x)$. Recalling the asymptotics of $G(x)$ at large $x$, the only choice is $g(x)={c}/x$ with some constant $c$. Let us fix the normalization of $Q(x)$ by choosing its free coefficient to be $1$, i.e. $Q(x)=ax^2+bx+1$. Then the remaining constants $a,b,c$ are fixed from finiteness of $G(x)$ at $x=0$ together with \eq{Gas}, \eq{Gat0}. This gives
\beq
\label{Gfin}
	G(x)=\ofrac{2}\(\ofrac{x}-\pi n\)+\ofrac{2x}\sqrt{(\pi n x)^2+(4\pi m-2\pi n)x+1}
\eeq
and
\beq
	\alpha=m/n,\ 
\eeq
so we must have $m<n$.

\begin{figure}[ht]
\begin{center}
\includegraphics[scale=0.7]{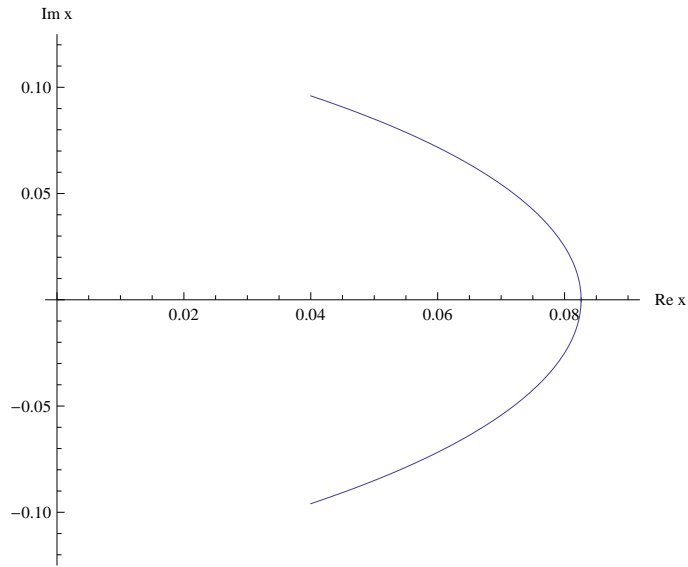}
\end{center}
\caption{ The branch cut of the resolvent \eq{Gfin} connecting the two branch points (plot generated from numerical solution of Bethe equations). Notice the bending of the cut. }
\label{fig:cut}
\end{figure}

It's important to understand a subtlety with the choice of branches of the square root in \eq{GfgQ}. Naively one might think that at $x=0$ the square root in \eq{Gfin} will be equal to $\sqrt{1}=+1$, but then the pole at $x=0$ wouldn't cancel. To clarify this let us consider as an example $n=3,m=1$. Then the branch points are at
\beq
	x_{1}=\frac{1+ 2i\sqrt{2}}{9\pi},\ \ x_{2}=\frac{1- 2i\sqrt{2}}{9\pi}
\eeq
and the cut should connect them. First, strictly speaking the cut is not a straight line in this case. If we take a small part of the cut approximated by the segment $\Delta x\in \mC$, then $L\rho(x)\Delta x$ is the number of Bethe roots inside this part of the cut. This quantity should be real, so the condition $\rho(x)dx\in\mR$ determines how exactly the cut will bend. Qualitatively it is shown on Fig. \ref{fig:cut}. Second, far away from the cut, in particular for large positive $x$,  we should have $\sqrt{Q(x)}\sim \pi n x>0$.
However if we start from large positive $x$ and go along the real axis to the point $x=0$, we will cross the cut. This means that when evaluating $G(x)$ at $x=0$ we will have to take $\sqrt{Q(x)}=-1$, not $+1$. Now it's clear that \eq{Gfin} indeed satisfies all the constraints we discussed\footnote{In {\texttt{Wolfram Mathematica}} the proper choice of branch cut in the square root $\sqrt{(\pi n x)^2+(4\pi m-2\pi n)x+1}$ appearing in \eq{Gfin} would be given for $n=3,m=1$ by
$3\pi i\sqrt{-i(x-x_1)}\sqrt{-i(x-x_2)}$ (this expression however doesn't take the bending of the cut into account). The extra factors of $i$ ensure that both square root factors have cuts going off to infinity in the same direction, so in their product the cut will disappear except between the branch points.}.

Finally, to extract the energy, notice that in the classical limit
\beq
	E=2\sum_j\frac{2}{w_j^2+1}\simeq\ofrac{L^2}\sum_j\ofrac{x_j^2}
	=\frac{2}{L}\int_C\frac{d\xi\rho(\xi)}{\xi^2}\ ,
\eeq
i.e. the energy is proportional to the linear coefficient of the resolvent's Taylor expansion in \eq{Gsmallx}.
From \eq{Gfin} we thus get
\beq
	E=\frac{2\pi^2m(n-m)}{L}\ .
\eeq
We see that we have found the energy by just imposing the correct analyticity, and not solving the Bethe equations directly at all! 

The spectral curve in this case consists of two Riemann sheets joined by a single cut, i.e. it is a sphere. Due to this the solution we found is known as a rational solution. The solution for the case with e.g. two branch cuts, corresponding to a torus, would be called an elliptic solution.

Let us finally note that the finite-gap equations played an important role in the devolpment of integrability in the AdS/CFT context. In particular, the classical limit of Bethe equations derived from gauge theory (conceptually similar to the discussion above) matches the classical spectral curve of the $AdS_5\times S^5$ integrable string sigma model, providing a nice demonstration of the AdS/CFT duality (see \cite{Beisert:2005di} and the review
\cite{SchaferNameki:2010jy}).

\section{Bethe ansatz for the SU(3) chiral Gross-Neveu model}
\label{sec:su3}

Now our goal is to study the $SU(3)$ chiral Gross-Neveu model. There are now more particle types, and we will first discuss only the particles in the fundamental representation of $SU(3)$. That is, we have three particle flavors and the flavor space for each particle is $\mC^3$. Our goal is to derive Bethe equations for these excitations. As other particles are their bound states, these Bethe equations are in fact enough to decribe the full spectrum (see e.g. \cite{Volin}).

\subsection{Nested Bethe ansatz for the SU(3) chain}

The method we will use is a more advanced version of the algebraic Bethe ansatz from section \ref{sec:su2gn}, known as the nested algebraic Bethe ansatz. \footnote{Let us note that the $SU(3)$ spin chain and many other models with higher rank symmetry group can also be solved by a \textit{coordinate} rather than algebraic version of the nested Bethe ansatz which also proved useful in AdS/CFT (see e.g. \cite{Beisert:2005tm} and references therien).}.

It will be more convenient to use a slightly different parameterization of the energy and momentum compared to what we had (Eq. \eq{pEsu2}) in the $SU(2)$ case, namely we take\footnote{We use this notation so that the R-matrix in \eq{Rsu3} that we get here  is the same as we had before in the $SU(2)$ case.}
\beq
\label{pEsu3}
	E=m\cosh \frac{\pi u}{3}, \ p=m\sinh \frac{\pi u}{3}\ .
\eeq
Then extracting from the S-matrix the scalar prefactor (see e.g. \cite{Volin} and references therein}),
\beq
	\hat S_{12}(p_1,p_2)=S^{su(3)}(u_1-u_2)\hat R_{12}^{-1}(u_1-u_2)
\eeq
\beq
	S^{su(3)}(u)=-\frac{\Gamma(1-\frac{u}{6i})\Gamma(\frac{2}{3}+\frac{u}{6i})}
	{\Gamma(1+\frac{u}{6i})\Gamma(\frac{2}{3}-\frac{u}{6i})}\ ,
\eeq
we have the R-matrix
\beq
\label{Rsu3}
	\hat R(u)=\ofrac{u+2i}(u+2i \hat P)\ ,
\eeq
which has the same form as in the $SU(2)$ case but now acts in $\mC^3\otimes\mC^3$. The main goal is to diagonalize the transfer matrix
\beq
	\hat T(u)=\Tr_a \hat T_a(u)=\Tr_a \(\hat R_{1a}(u_1-u)\hat R_{2a}(u_2-u)\dots \hat R_{na}(u_n-u) \)\ .
\eeq
Let us write out its structure in the auxiliary space explicitly, in the following notation:
\beq 
	\hat T_a(u)=
	\begin{pmatrix}
  \hat T_{00}(u) & \hat B_1(u) & \hat B_2(u) \\
  \hat C_1(u) & \hat T_{11}(u) & \hat T_{12}(u) \\
\hat C_2(u) & \hat T_{21}(u) & \hat T_{22}(u)
 \end{pmatrix}\ .
\eeq
The commutation relations between the entries follow from the RTT relation as usual. In particular,
\beq
	[\hat B_1(u),\hat B_1(v)]=0,\ \ \ \ \ 	[\hat B_2(u),\hat B_2(v)]=0\ ,
\eeq
\beq
	\hat B_1(u)\hat B_2(v)=\frac{v-u}{v-u+2i}\hat B_2(v)\hat B_1(u)+\frac{2i}{v-u+2i}\hat B_1(v)\hat B_2(u)\ .
\eeq
While one can do the calculation in a more abstract way we will use index notation to ensure full clarity. 
We will have Greek indices $\alpha,\beta,\dots$ and Latin indices $a,b,\dots$, all of which take values 1 and 2. In this notation the relations for commuting the $\hat T$'s with the $\hat B$'s read
\beq
	\hat T_{00}(u)\hat B_\alpha(v)=\frac{v-u-2i}{v-u}\hat B_\alpha(v)\hat T_{00}(u)
	{\color{red}+\frac{2i}{v-u}\hat B_\alpha(u)\hat T_{00}(v)}
\eeq
and
\beq
	\hat T_{\alpha\alpha'}(u)\hat B_\beta(v)=\frac{v-u+2i}{v-u}{\mathbb R}^{\tau\gamma}_{\alpha'\beta}(v-u)\hat B_\gamma(v)\hat T_{\alpha\tau}(u)
	{\color{red}+{\frac{2i}{u-v}\hat B_{\alpha'}(u)\hat T_{\alpha\beta}(v)}}\ ,
\eeq
where we marked by red the 'unwanted' terms, i.e. those that later will cancel when we construct the eigenstate and impose the Bethe equations. Remarkably, the $SU(2)$ R-matrix which we denote as ${\mathbb R}(u)$, appears in these equations. Explicitly its nonzero elements are, as before,
\beq
	{\mathbb R}_{11}^{11}(u)={\mathbb R}_{22}^{22}(u)=1, \ 
\eeq
\beq
	{\mathbb R}_{21}^{21}(u)={\mathbb R}_{12}^{12}(u)=\frac{u}{u+2i},\ \ \ \ 
	{\mathbb R}_{12}^{21}(u)={\mathbb R}_{21}^{12}(u)=\frac{2i}{u+2i} \ 
\eeq

While in the $SU(2)$ case we had $\hat B(u)$ as a creation operator, here we have two candidates -- $\hat B_1(u)$ and $\hat B_2(u)$. Let us try to build the eigenvectors as
\beq
\label{PBsu3}
	|\Psi\rangle=\sum_{\{a\}}\hat B_{a_1}(v_1)\hat B_{a_2}(v_2)\dots \hat B_{a_n}(v_n)F^{a_1a_2\dots a_n} |0\rangle
\eeq
with $a_1,a_2,\dots$ taking the values $1$ or $2$, and the vacuum is as usual
\beq
	\ket{0}=\ket{\uparrow\uparrow\dots \uparrow}=
	\begin{pmatrix}1\\0\\0\end{pmatrix}\otimes\dots\otimes\begin{pmatrix}1\\0\\0\end{pmatrix}\ .
\eeq
Let's assume for now that there are no unwanted (i.e. red) terms in the commutation relations, and act with $T(u)$ on this state. From $\hat T_{00}$ we get
\beq
	\hat T_{00}(u)|0\rangle=|0\rangle
\eeq
and 
\beq
	\hat T_{00}(u)|\Psi\rangle=\prod_k\frac{v_k-u-2i}{v_k-u}|\Psi\rangle\ .
\eeq
Then let's see what we get acting on our state $|\Psi\rangle$ with the rest of the trace of $T(u)$, i.e. with $\sum_\alpha \hat T_{\alpha\alpha}$. Let's take for a start a state $|\Psi\rangle$ with only two excitations,
\beq
	|\Psi\rangle=\hat B_{a_1}(v_1)\hat B_{a_2}(v_2)F^{a_1a_2}|0\rangle\ .
\eeq
Then
\beqa
	\hat T_{\alpha\alpha}(u)|\Psi\rangle &=&
	\frac{v_1-u+2i}{v_1-u}{\mR}_{\alpha a_1}^{\tau_1b_1}(v_1-u)\hat B_{b_1}(v_1)\hat T_{\alpha\tau_1}(u)
	\hat B_{a_2}(v_2)F^{a_1a_2}|0\rangle
	\\ \nn
	&=&
	\frac{v_1-u+2i}{v_1-u}\frac{v_2-u+2i}{v_2-u}
	{\mR}_{\alpha a_1}^{\tau_1b_1}(v_1-u){\mR}_{\tau_1 a_2}^{\tau_2b_2}(v_2-u)
	\\ \nn
	&\times&
	\hat B_{b_1}(v_1)\hat B_{b_2}(v_2)\hat T_{\alpha\tau_2}(u)
	F^{a_1a_2}|0\rangle\ .
\eeqa
Finally, the operators $\hat T_{\alpha\beta}$ act on the vacuum in a very simple way. That is,
\beq
	\hat T_{\alpha\beta}(u)|0\rangle=\delta_{\alpha\beta}\prod_{j}\frac{u_j-u}{u_j-u+2i}\;|0\rangle\ .
\eeq
So we get
\beqa
	\hat T_{\alpha\alpha}(u)|\Psi\rangle &=&
	\frac{v_1-u+2i}{v_1-u}\frac{v_2-u+2i}{v_2-u}\prod_{j}\frac{u_j-u}{u_j-u+2i}
	\\ \nn
	&\times&
	{\mR}_{\tau_2 a_1}^{\tau_1b_1}(v_1-u){\mR}_{\tau_1 a_2}^{\tau_2b_2}(v_2-u)
	\hat B_{b_1}(v_1)\hat B_{b_2}(v_2)
	F^{a_1a_2}|0\rangle \ . 
\eeqa
In the indices of the $\mR$-matrices there is now a clear pattern, so we see that for any number of excitations as in \eq{PBsu3} we would get
\beqa
	\hat T_{\alpha\alpha}(u)|\Psi\rangle &=& \prod_{k=1}^n\frac{v_k-u+2i}{v_k-u}\prod_{j}\frac{u_j-u}{u_j-u+2i}
	\prod_{k=1}^n \hat B_{b_k}(v_k)|0\rangle 
	\\ \nn
	&\times&
		{\mR}_{\tau_2 a_1}^{\tau_1b_1}(v_1-u) {\mR}_{\tau_3 a_2}^{\tau_2b_2}(v_2-u)
		\dots
		{\mR}_{\tau_{1} a_n}^{\tau_nb_n}(v_n-u)
	F^{a_1a_2\dots a_n}\ .
\eeqa
But this product of $\mR$-matrices is the transfer matrix of an $SU(2)$ spin chain! It has free indices $a_1,\dots,a_n$ and $b_1,\dots,b_n$, so it acts on the product of vector spaces $V_1\otimes\dots\otimes V_n$ with each $V_j\simeq\mC^2$. And $F^{a_1a_2\dots a_n}$ is the set of coordinates of a vector in that tensor product, or in other words the wavefunction of an $SU(2)$ chain with $n$ sites. The first upper index and the first lower index of each R-matrix are the indices in the auxiliary space and they are contracted with the neighbouring R-matrices. So we have a product of R-matrices along the auxiliary space and we take a trace over this space. This is precisely the transfer matrix of the $SU(2)$ spin chain on $n$ sites. Notice that $v_1,\dots,v_n$ are the \textit{inhomogenieties} in this chain. 

The $a_k$ indices are contracted with $F$, so if $F$ is an eigenstate of this transfer matrix we see that $T_{\alpha\alpha}(u)$ will act diagonally on $|\Psi\rangle$ -- which is what we want. So we have reduced the initial $SU(3)$ problem to an $SU(2)$ one. This is why the approach we are discussing is called ``\textit{nested} algebraic Bethe ansatz''.

Of course we already solved the $SU(2)$ spin chain, so we know how to diagonalize this transfer matrix. States will be created by its off-diagonal element, and are parameterized by yet another set of Bethe roots, which we call $w_m$. (It's the third one -- in addition to $u_j$ and $v_k$.) Then we take as $F^{a_1\dots a_n}$ the wavefunction of this state:
\beq
	F^{a_1\dots a_n}=\(| {\rm{\ SU(2) \ eigenstate\ }}\rangle\)^{a_1\dots a_n}\ .
\eeq
For example, $F^{122}$ is the coefficient of the term $\ket{\upa\dna\dna}$ in the eigenstate of the $SU(2)$ chain. 
The new Bethe roots satisfy the $SU(2)$ Bethe equations \eq{abasu21} with inhomogenieties set to be $v_k$:
\beq
\label{baensu2}
	\prod_k\frac{w_m-v_k+i}{w_m-v_k-i}=\prod_{m'\neq m}\frac{w_m-w_{m'}+2i}{w_m-w_{m'}-2i}\ .
\eeq
The corresponding eigenvalue of the $SU(2)$ transfer matrix is given by \eq{lamsu21},
\beq
\label{lamnsu2}
	\Lambda_{SU(2)}(u)=\prod_m\frac{u-w_m+i}{u-w_m-i}+
	\prod_k\frac{u-v_k}{u-v_k-2i}\prod_m\frac{u-w_m-3i}{u-w_m-i}\ .
\eeq

Then, we can assemble all our calculations to write the result for the full transfer matrix eigenvalue,
\beq
	(\hat T_{00}(u)+\hat T_{11}(u)+\hat T_{22}(u))|0\rangle =\Lambda(u)|0\rangle 
\eeq
where
\beqa
	\Lambda(u)&=&
	\prod_k\frac{v_k-u-2i}{v_k-u}+
	\Lambda_{SU(2)}(u)
	\prod_k\frac{v_k-u+2i}{v_k-u}\prod_j\frac{u_j-u}{u_j-u+2i}\ .
\eeqa

During this whole discussion we of course ignored the extra terms in the commutation relations. One can show that they will cancel provided the roots $v_k$ satisfy a set of constraints which are the Bethe equations for this model. Showing this requires some careful and lengthy work, and we will not do this here.
Some guidance for a similar model can be found in \cite{Essler:1992he}.

Rather than going into details of this derivation we will follow instead a shortcut which gives the same Bethe equations. Recall that for $SU(2)$ one way to derive the Bethe equations is to require cancellation of spurious poles in the transfer matrix eigenvalue. We can do the same here. We see that $\Lambda(u)$ appears to have poles when $u=v_k$ which in fact should be absent as the transfer matrix is not singular at those points. Demanding that the residue at $u=v_{k'}$ vanishes we find
\beq
	0=(-2i)\prod_{k\neq k'}\frac{v_k-v_{k'}-2i}{v_k-v_{k'}}+
	\Lambda_{SU(2)}(v_{k'})\times 2i \times
	\prod_{k\neq k'}\frac{v_k-v_{k'}+2i}{v_k-v_{k'}}\prod_j\frac{u_j-v_{k'}}{u_j-v_{k'}+2i}
\eeq
(note that $\Lambda_{SU(2)}$ is clearly not singular at this point). Plugging in $\Lambda_{SU(2)}(v_{k'})$ from \eq{lamnsu2}, where only the first term is nonzero, we get
\beq
	0=-\prod_{k\neq k'}\frac{v_k-v_{k'}-2i}{v_k-v_{k'}+2i}+\prod_{m}\frac{v_{k'}-w_m+i}{v_{k'}-w_m-i}
	\prod_j\frac{u_j-v_{k'}}{u_j-v_{k'}+2i}\ .
\eeq
Shifting $v_k=\tilde v_k+i,w_m=\tilde w_m+i$ and dropping the tildes we get
\beq
	0=-\prod_{k\neq k'}\frac{v_k-v_{k'}-2i}{v_k-v_{k'}+2i}+\prod_{m}\frac{v_{k'}-w_m+i}{v_{k'}-w_m-i}
	\prod_j\frac{u_j-v_{k'}-i}{u_j-v_{k'}+i}
\eeq
Note that the Bethe equations \eq{baensu2} for $w$'s do not change under this shifting.
So, we find
\beq
	\prod_j\frac{v_{k'}-u_j+i}{v_{k'}-u_j-i}=
	\prod_{k\neq k'}\frac{v_{k'}-v_{k}+2i}{v_{k'}-v_{k}-2i}
	\prod_{m}\frac{v_{k'}-w_m-i}{v_{k'}-w_m+i}\ ,
\eeq
while the $w$'s satisfy the same equations as before in \eq{abasu21} but with inhomogeneities $v_k$,
\beq
	\prod_{m}\frac{w_j-u_m+i}{w_j-u_m-i}=\prod_{k\neq j}\frac{w_j-w_k+2i}{w_j-w_k-2i} \ .
\eeq

The equations above determine the eigenvalue of the $SU(3)$ transfer matrix and in particular of the $SU(3)$ version of the XXX spin chain Hamiltonian. Like in the $SU(2)$ case this Hamiltonian is expressed in terms of the transfer matrix as in \eq{HXXXT}
\beq
	\hat H=\left. i\frac{d}{du}\log \hat T(u)\right|_{u=0}\ ,
\eeq
and can be written explicitly as 
\beq
	\hat H={\rm{const}} \times\sum_{k=1}^{N_f}\(1-\hat P_{k,k+1}\)\ ,
\eeq
the only difference with the $SU(2)$ case being that it now acts in the tensor product of $\mC^3$ rather than $\mC^2$ spaces.

Let us also note that for the $SU(2)$ chain we had one type of Bethe roots ($w_k$) parameterizing the eigenstates, while here we have two types -- $v_k$ and $w_k$. In fact one should think of each Bethe root type as associated to the nodes on the Dynkin diagram of the symmetry group. In our case, accordingly, $SU(2)$ has a Dynkin diagram with only one node, while for $SU(3)$ there are two nodes. One can also consider spin chains with higher rank symmetry groups $G$ beyond the $SU(2)$ and $SU(3)$ cases. Accordingly, instead of $\mC^2$ or $\mC^3$ one one would have at each site of the chain a higher-dimensional complex space. At least for most of the compact simple Lie groups $G$, the corresponding spin chain again can be solved by Bethe ansatz, and the Bethe equations are written in a uniform way in terms of the group's Cartan matrix as well the representation of the group chosen at each site (see e.g. the review \cite{Volin} and references therein). A similar story continues to hold even for super Lie algebras, in particular for the algebra $\mathfrak{psu}(2,2|4)$ which underlies the structure of Bethe equations describing the spectrum of long operators in $\cN=4$ supersymmetric Yang-Mills theory \cite{Beisert:2010jr}.

Finally we can assemble the equations for the spectrum of the $SU(3)$ chiral Gross-Neveu model. Using the expression above for the eigenvalue of the $SU(3)$ transfer matrix, we can write the periodicity condition as
\beq
	e^{ip_jL}\prod_mS^{su(3)}(u_j-u_m)\prod_k\frac{u_j-v_k+i}{u_j-v_k-i}=-1\ ,
\eeq
together with equations for auxiliary Bethe roots
\beq
	\prod_j\frac{v_{k'}-u_j+i}{v_{k'}-u_j-i}=
	\prod_{k\neq k'}\frac{v_{k'}-v_{k}+2i}{v_{k'}-v_{k}-2i}
	\prod_{m}\frac{v_{k'}-w_m-i}{v_{k'}-w_m+i}\ ,
\eeq
\beq
	\prod_{m}\frac{w_j-u_m+i}{w_j-u_m-i}=\prod_{k\neq j}\frac{w_j-w_k+2i}{w_j-w_k-2i}\ .
\eeq
The energies are as usual a sum of single particle energies (notice we use the parameterization \eq{pEsu3})
\beq
	E=\sum_j m \cosh \frac{\pi u_j}{3}\ .
\eeq
Thus we have completed the solution for the spectrum of the $SU(3)$ Gross-Neveu model in large volume $L$.

%
%
%


\section{Bethe ansatz for the harmonic oscillator}
\label{sec:osc}

Let us discuss in this last section a completely different setting where Bethe-like equations also appear. Namely, one can use a kind of Bethe ansatz to get eigenstates of the very well-studied one-dimensional harmonic oscillator in quantum mechanics (for a more detailed discussion of this case see e.g. \cite{Gromov:2007aq}). Thus, we are studying the Schrodinger equation
\beq
	-\frac{\hbar^2}{2m}\psi''(x)+V(x)\psi(x)=E\psi(x)
\eeq
with the potential
\beq
	V(x)=\frac{m\omega^2x^2}{2}\ .
\eeq
Let us introduce the so-called quasimomentum
\beq
\label{ppsi}
	p(x)=\frac{\hbar}{i}\frac{\psi'(x)}{\psi(x)}\ ,
\eeq
in terms of which the Schrodinger equation takes the form
\beq
\label{peq}
	p^2-i\hbar p'=2m(E-V)\ .
\eeq
As $\psi(x)$ is regular, the only singularities of $p(x)$ are at the zeros $x=x_j$ of the wavefunction, where the quasimomentum has simple poles with residue $\frac{\hbar}{i}$.

In the classical limit, i.e. for highly excited states, we get from \eq{peq}
\beq
	p\simeq p_{cl}=\sqrt{2m(E-V)}
\eeq
so now the quasimomentum has a branch cut. This cut can be understood as a collection of poles at $x_j$, which become denser and denser, eventually forming a smooth distribution giving rise to a cut. The situation is very similar to the classical spectral curve of the XXX model we discussed in section \ref{sec:curve}

Let us now see that for any state (not necessarily semiclassical) we can derive a simple set of equations fixing the positions of these poles. At large $x$ we have
\beq
	p(x)=im\omega x+\cO(1/x)\ ,
\eeq
so we can write
\beq
\label{pexplic}
	p(x)=im\omega x+\frac{\hbar}{i}\sum_{j=1}^N\frac{1}{x-x_j}\ .
\eeq
From the large $x$ asymptotics of \eq{peq} we can already find the spectrum! It reads
\beq
	E=\hbar \omega\(N+\frac{1}{2}\)\ .
\eeq
From \eq{peq} we also get a set of Bethe-like equations for the roots,
\beq
\label{harmba}
	x_j=\frac{\hbar}{2\omega m}\sum_{k \neq j}\frac{1}{x_j-x_k}\ .
\eeq
They are clearly reminiscent of the usual Bethe ansatz form, with one root in the l.h.s. and interaction between roots in the r.h.s. As one can expect from the usual form of the oscillator wavefunctions, the solutions to this equation are the roots of the $N$-th Hermite polynomial,
\beq
	H_N\(\sqrt{\frac{2m\omega}{\hbar}}x_j\)=0\ .
\eeq
Knowing the roots $x_j$ we can also reconstruct the wavefunction from \eq{ppsi}, \eq{pexplic}. 
Equations similar to \eq{harmba} frequently arise as limiting cases of the Bethe ansatz equations for other models, e.g. in the limit of large $L$ and fixed number of excitations in the XXX spin chain.



\section*{Acknowledgements}

This article arose out of a series of lectures gven by the author at the Young Researchers Integrability School at Durham University in July 2015. I wish to thank all the organizers and the other lecturers for their effort in making the event run successfully and creating a great atmosphere. I would like to also thank the Department of Mathematical Sciences at Durham University for hospitality.  I am particularly grateful to Z. Bajnok, D. Bombardelli, N. Gromov, A. Sfondrini, S. van Tongeren and A. Torrielli for discussions related to the material in these notes. I am also grateful to the student participants for their interest in the school, excellent questions and lots of feedback. My work is supported in part by funding from the People Programme (Marie Curie Actions) of the European Union's Seventh Framework Programme
FP7/2007-2013/ under REA Grant Agreement No 317089 (GATIS).

\end{document}